\documentclass{article}
\usepackage{scalefnt}
\usepackage{algorithm}
\usepackage{algpseudocode}
\usepackage{hyperref}
 \usepackage{booktabs}
\usepackage{spconf2,amsmath,graphicx,amssymb,bm,bbm}
\usepackage{dsfont}
\usepackage{hyperref, url}
\usepackage{pgf-pie}
 \usepackage{booktabs}
 \usepackage{multirow}
 \usepackage{caption}
\usepackage{subcaption}
 \usepackage[english]{babel}
 \usepackage[utf8]{inputenc}
 \usepackage[T1]{fontenc}
 \usepackage{lipsum}
\newcommand\thefont{\expandafter\string\the\font}

\title{Solving Audio Inverse Problems with a Diffusion Model}
 \name{Eloi Moliner$^1$ \qquad Jaakko Lehtinen$^2$ \qquad Vesa Välimäki$^1$\sthanks{This research is part of the activities of the Nordic Sound and Music Computing Network---NordicSMC (NordForsk project no.~86892).}}
  
  \address{$^1$ Acoustics Lab, Dept. of Information \& Communications Eng. \quad $^2$ Dept. of Computer Science\\
	Aalto University, Espoo, Finland\\
    eloi.moliner@aalto.fi}
\begin{document}
\ninept
\maketitle
\begin{abstract}
This paper presents CQT-Diff, a data-driven generative audio model that can, once trained, be used for solving various different audio inverse problems in a problem-agnostic setting. CQT-Diff is a neural diffusion model with an architecture that is carefully constructed to exploit pitch-equivariant symmetries in music. This is achieved by preconditioning the model with an invertible Constant-Q Transform (CQT), whose logarithmically-spaced frequency axis represents pitch equivariance as translation equivariance.
The proposed method is evaluated with solo piano music, using objective and subjective metrics in three different and varied tasks: audio bandwidth extension, inpainting, and declipping. The results show that CQT-Diff outperforms the compared baselines and ablations in audio bandwidth extension and, without retraining, delivers competitive performance against modern baselines in audio inpainting and declipping. 
This work represents the first diffusion-based  general framework for solving inverse problems in audio processing.





\end{abstract}
\begin{keywords}Audio systems, deep learning, inverse problems, signal restoration.
\end{keywords}
%
\section{Introduction}
\label{sec:intro}

Audio restoration tasks, such as bandwidth extension, inpainting, and declipping, are inverse problems with the aim to restore the signal from observations that have suffered a known type of degradation. These problems are ill-posed in the sense that several different restorations may be equally plausible.
Algorithms are typically constructed for a particular type of degradation using domain knowledge, such as signal sparsity \cite{kitic2015sparsity, siedenburg2014audio}, or the low-rankness of the power spectrum \cite{bilen2018solving}. Data-driven generative models have also been recently proposed \cite{moliner2022behm, marafioti2020gacela}. 
A common shortcoming is that an algorithm engineered for one type of degradation is typically not useful for others.


In this work, we explore whether a single pre-trained generative audio model could be useful for different restoration tasks \emph{without knowledge of which types of degradations it will be applied to at inference time}. We base our work on diffusion models \cite{ho2020denoising, song2020score}, a family of generative models that have shown outstanding performance in different modalities, such as image 
\cite{rombach2022high}, 
video \cite{ho2022video}, speech  \cite{kong2021diffwave,  serra2022universal}, and audio generation \cite{rouard2021crash, hawthorne2022multi}.
Particularly relevant to us is that unconditional (purely generative) diffusion models have proven useful also in conditional restoration settings where parts of the result are pre-determined, particularly in the image domain \cite{kawar2022denoising,  chung2022improving, chung2022diffusion}.
Crucially, as these approaches do not require paired data for training and the degradation model is only used during inference, the models adapt to a new problem without retraining. To the best of our knowledge, analogous methods have not been employed in audio audio restoration. This paper addresses this research gap.

Training an unconditional audio diffusion model is the first step towards this objective.
We argue that this is overly challenging unless suitable inductive biases are built in to the model. While unconditional audio generation has been demonstrated using a raw audio waveform representation, the experiments are limited to small speech datasets \cite{kong2021diffwave, goel2022sashimi}, or non-tonal audio  \cite{rouard2021crash, Moliner2022realistic}, with performance expected to decrease for more challenging datasets.
Building diffusion models using mel-spectrograms 
or magnitude spectrograms \cite{hawthorne2022multi} leads to a more expressive generative performance. However, the lack of invertibility of these transforms blocks the required conditioning for solving inverse problems. We design our diffusion model in the time domain, allowing for maximum flexibility, and use an invertible time-frequency transform as a preprocessing step.
    
Choosing the right transform plays a critical role in the performance of the model. 
One option is to use a Short-Time Fourier Transform (STFT) and apply a neural network based on 2-D convolutions, adapted from image generation.
While this approach succeeded for speech \cite{Richter2022Speech}, we argue that, contrary to image processing, it is non-ideal for signals with strong harmonics (music), as translation equivariance of harmonics does not hold true in linear frequency. 
The Constant-Q Transform (CQT), in which the frequency axis is logarithmically spaced, is a possible solution to this problem. Since pitch-shifting corresponds to translation in the CQT spectrogram, applying the convolutional operator is now appropriate.
The CQT is widely used in Music Information Retrieval \cite{won2020data}, but, despite some exceptions \cite{huang2018timbretron}, is not common in audio generation tasks.

We first show that CQT preconditioning yields a high-quality diffusion model for audio. We present a versatile framework called CQT-Diff for solving audio restoration problems, from which we evaluate three: bandwidth extension, audio inpainting, and declipping. Pitted against specialized algorithms, our model yields state-of-the art or strongly competitive performance in listening tests.

\section{CQT-Diff framework}

\subsection{Score-based
Generative Modeling}
Consider a diffusion process where data $\bm{x}_0 \sim p_\text{data}$ is progressively diffused into Gaussian noise $\bm{x}_{\tau_\text{max}} \sim \mathcal{N}(\mathbf{0},\sigma_\text{max}^2 \mathbf{I})$ over time\footnote{The ``diffusion time'' $\tau$ must not be confused with the ``audio time'' $t$. 
} $\tau$ by applying a perturbation kernel $p_t(\bm{x}_\tau)=\mathcal{N}(\bm{x}_\tau;\bm{x}_0,\sigma_\tau^2 \mathbf{I})$.
Diffusion models, also known as score-based models, generate data samples by reversing the aforementioned diffusion process. 
Specifically, diffusion models estimate the gradient of the log probability density with respect to data $\nabla_{\bm{x}_\tau}\log p_\tau(\bm{x}_\tau)$, known as the \textit{score} function. 

Following \cite{karras2022elucidating}, the reverse diffusion process can be defined by the following Stochastic Differential Equation (SDE):
\begin{equation}\label{sde}
    \text{d}\bm{x} =-\sigma_\tau(\beta(\tau)\sigma_\tau+\dot\sigma_\tau)\;\nabla_{\bm{x}}\log p_\tau(\bm{x}_\tau) \text{d}\tau +\sqrt{2\beta(\tau)}\sigma_\tau\;d\text{w},
\end{equation}
where d$\tau$ is an infinitesimal negative timestep, $d\text{w}$ is the standard Wiener process when time flows backwards, $\dot\sigma_\tau$ is the time derivative of the noise variance schedule, and $\beta(\tau)$ is, intuitively, a function that defines the amount of stochasticity added into the process.
Note that stochasticity is, in fact, not needed and \eqref{sde} could be defined as an ordinary differential equation \cite{song2020score, karras2022elucidating}. However, stochasticity has a role in correcting approximation errors during sampling \cite{karras2022elucidating} and is empirically found to improve the generation quality in our model.


The score is intractable but can be approximated with a deep neural network $s_\theta(\bm{x}_\tau, \sigma_\tau) \simeq \nabla_{\bm{x}_\tau}\log p_\tau(\bm{x}_\tau; \sigma_\tau)$.
The parameterized score can be written as
$    s_\theta(\bm{x}_\tau,\sigma_\tau) =(D_\theta(\bm{x}_\tau,\sigma_\tau)-\bm{x}_\tau)/\sigma_\tau^2,$
 where  $D_\theta(\bm{x}_\tau,\sigma_\tau)$ is a denoiser to be trained with the L2 objective:
\begin{equation}\label{loss}
    \mathbb{E}_{\bm{x_0} \sim p_\text{data}, \boldsymbol\epsilon \sim \mathcal{N}(\mathbf{0},\mathbf{I}) }  \left[ \lambda(\sigma_\tau) \lVert D_\theta(\bm{x_0}+\sigma_\tau\bm{\epsilon},\sigma_\tau) -\bm{x}   \rVert_2^2 \right],
\end{equation}
where $\lambda(\sigma_\tau)$ is a weighting function.
Once the model has been trained, one can generate new data instances by substituting the estimated score $s_\theta(\bm{x}_\tau, \sigma_\tau)$ in \eqref{sde}, and solving the SDE with a numerical solver of choice.

\subsection{Diffusion Models for Inverse Problems}
Consider the goal of retrieving an unknown audio signal $\bm{x_0}$ from a set of measurements or observations $\bm{y} =\mathcal{A}(\bm{x_0}) + \bm{\epsilon}$, where $\mathcal{A}(\cdot)$ is a known and differentiable degradation function, and $\bm{\epsilon} \sim \mathcal{N}(0,\sigma_y^2 \bm{I})$ is measurement noise. The objective would be to sample from the posterior distribution given the observations $p(\bm{x} | \bm{y})$. In the context of a diffusion model,  this would require substituting the score in \eqref{sde} for the conditional score $ \nabla_{\bm{x}_\tau}\log p_\tau(\bm{x}_\tau|\bm{y})$.

One solution is to train a score model $s_\theta(\bm{x}_\tau, \bm{y}, \tau)$ that depends explicitly on the observations \cite{ han2022nu, serra2022universal, rombach2022high}. However, these approaches lack versatility as they require training with paired data and cannot effectively adapt to unseen degradations.
 Instead, applying Bayes, the posterior factorizes as $p(\bm{x}|\bm{y}) \propto p(\bm{x}) p(\bm{y|x})$, which leads to
 \begin{equation}\label{posterior_score}
    \nabla_{\bm{x}_\tau}\log p_\tau(\bm{x}_\tau|\bm{y})= \nabla_{\bm{x}}\log p_\tau(\bm{x}_\tau)+
     \nabla_{\bm{x}_\tau}\log p_\tau(\bm{y}|\bm{x}_\tau).
\end{equation}
Then, an unconditional diffusion model can be used as a generative prior, while the conditioning is applied by some other means.
 In some cases, the likelihood  can be indirectly applied by ensuring data consistency during sampling, replacing parts of the internal predictions $\hat{\bm{x}}_0=D_\theta(\bm{x}_\tau,\sigma_\tau)$ with the measurements $\bm{y}$ during every discretized step \cite{song2020score, kawar2022denoising, choi2021ilvr, lugmayr2022repaint}. 
 We refer to this strategy as the \textit{data consistency} (DC) method.
Note that this method can only be used in cases where an explicit replacement is available. Degradations such as distortions or nonlinear-phase filtering cannot be treated by replacing parts of the signal in the time domain.
 
Recent works \cite{chung2022improving, chung2022diffusion,ho2022video} demonstrated that the aforementioned data consistency trick produces suboptimal results, so they proposed an alternative strategy defining the second term of \eqref{posterior_score} as
\begin{equation}\label{recguid}
    \nabla_{\bm{x}_\tau}\log p_\tau(\bm{y}|\bm{x}_\tau) \simeq
    -\xi(\tau) \; \nabla_{\bm{x}_\tau} \lVert \bm{y} - \mathcal{A}(D_\theta(\bm{x}_\tau,\sigma_\tau)) \rVert^2.
\end{equation}
This solution derives from approximating the likelihood as a Gaussian distribution of the form $p_t(\bm{y}|\bm{x}_\tau)\sim \mathcal{N}(\mathcal{A}(D_\theta(\bm{x}_\tau ,\sigma_\tau)), \sigma_y^2 \mathbf{I})$.
Following \cite{ho2022video}, we refer to this method as \textit{reconstruction guidance} (RG). The gradient operator $\nabla_{\bm{x}_\tau}$ requires backpropagating through the degradation function $\mathcal{A}(\cdot)$, and through the  neural network of the denoiser $D_\theta(\bm{x}_\tau,\sigma_\tau)$. The scaling function $\xi(\tau)$ defines the amount of guidance that is applied during the reverse diffusion process. Inspired by \cite{kim2022guided}, we empirically found that a robust parameterization of the scaling function is 
$\xi(\tau)=\xi^\prime \sqrt{N}/( \tau \lVert G \rVert^2)$,
 where $G=\nabla_{\bm{x}_\tau} \lVert y - \mathcal{A}(D_\theta(\bm{x}_\tau,\sigma_\tau)) \rVert^2$ are the guidance gradients and $N$ is the audio segment size in samples. Then, choosing $\xi^\prime=0$ would lead to an unconditional sampler, but selecting too large a value for $\xi^\prime$ would often result with degenerated solutions. 
Note that RG and DC can also be combined. However, if the observations are noisy and, as a consequence, unreliable, DC cannot be trivially applied.

\subsection{Data Representation and Preprocessing}

To minimize error propagation during sampling, we apply a similar preconditioning as Karras {\it et al.}~\cite{karras2022elucidating}. The denoiser is defined as
\begin{equation}
D_\theta(\bm{x}_\tau, \sigma_\tau)=
c_\text{skip}(\sigma_\tau)\bm{x}_\tau+
c_\text{out}(\sigma_\tau)F_\theta(c_\text{in}(\sigma_\tau)\bm{x}_\tau, \tfrac{\ln(\sigma_\tau)}{4}),
\end{equation}
where $c_\text{skip}(\sigma_\tau)$, $c_\text{out}(\sigma_\tau)$, and $c_\text{in}(\sigma_\tau)$ are scaling parameters optimized in such a way that the input and the output of $F_\theta$ have always close-to-unit variance\footnote{These parameters are defined in Table 1 in \cite{karras2022elucidating} and depend on the standard deviation of the dataset, which needs to be estimated beforehand.}. Following the same approach, the loss weighting in \eqref{loss} is defined as $\lambda(\sigma_\tau)=1/c_\text{out}(\sigma_\tau)^2$.

Throughout this paper, $\bm{x}$ and $\bm{y}$ refer to signals in the raw waveform domain. However, in our early exploration, we found that using the time-domain waveform representation resulted in unsatisfactory results, especially for music bandwidth extension, which requires preserving the harmonic structure.
Instead, the data is transformed using an invertible CQT (ICQT), which is applied as a data preprocessing step inside $F_\theta$.
As stated in Sec.~\ref{sec:intro}, the motivation for using the CQT is to explore the partial pitch-equivariance symmetry that audio data exhibits in the log-spaced frequency range.
Thus, $F_\theta$ is composed as
$F_\theta =\text{ICQT} \circ F_\theta^\prime \circ \text{CQT}$,
where $F_\theta^\prime$ refers to the raw neural network layers  operating in the transform domain. Using this approach, the gradient is backpropagated through the forward CQT and its inverse, or ICQT, and, thus, the chosen CQT implementation must be both invertible and differentiable.

We use the CQT implementation based on non-stationary Gabor frames  \cite{velasco2011constructing}, which guarantees perfect reconstruction and is differentiable, since it is a linear transform implemented by applying FFT-processing.
If set up for minimum redundancy, this transform yields a non-uniform sampling grid where the time resolution increases with frequency. Alternatively,  we use a rasterized CQT representation where the hop sizes for all frequencies are set to the minimum, hence keeping the time resolution constant. This solution allows CQT processing as a 2D matrix at the cost of adding redundancy to the data.
In our experiments, we use a frequency resolution of 64 bins per octave, and a span of seven octaves.
\begin{figure}[t]
    \centering
    \includegraphics[width=0.96\columnwidth]{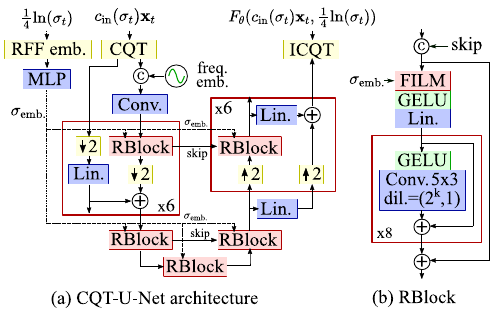}
    \caption{Diagram of the neural network architecture adapted for the CQT domain. 
    The design follows a U-Net structure, where only the time resolution is modified. Each intermediate \textit{RBlock} contains a stack of dilated convolutions in the frequency axis.
    }
    \label{fig:diagram}
\end{figure}

\subsection{Architecture Design}


Next, we summarize how CQT-Diff fits audio data in the CQT-domain using the neural network architecture shown in Fig.~\ref{fig:diagram}.
The architecture is a U-Net where only the time resolution is downscaled after each subsequent residual block (\textit{RBlock}), while the frequency resolution is maintained. The non-local dependencies in the frequency domain are tackled with stacks of exponentially-increasing dilated convolutions without bias, accompanied with GELU nonlinearities and residual connections. Inspired by \cite{song2020score}, we use anti-aliased downsampling operators and a progressive growing structure for the encoder and decoder sides \cite{karras2020analyzing}.
The diffusion step is shown to the network as $\tfrac{1}{4} \ln (\sigma_t)$, and is encoded using Random Fourier Features (RFF) \cite{tancik2020fourier} and a multi-layer perceptron. 
The embeddings are applied as a conditioner in every \textit{RBlock} using FiLM modulations \cite{perez2018film}. 
We concatenate RFF-based absolute frequency positional embeddings to the CQT-transformed input, meant to propagate absolute frequency information in the early layers of the network.




\section{Experiments and results}
This section summarizes our experiments on three different audio inverse problems: bandwidth extension, inpainting, and declipping.
Audio examples are included in the companion webpage\footnote{\href{http://research.spa.aalto.fi/publications/papers/icassp23-cqt-diff/}{research.spa.aalto.fi/publications/papers/icassp23-cqt-diff}}, 
and reproducible code for all the experiments can be found in the official repository\footnote{\href{https://github.com/eloimoliner/CQTdiff}{github.com/eloimoliner/CQTdiff}}, to which we refer for further implementation details. 


We train and test our model on the MAESTRO dataset \cite{hawthorne2018enabling}, which contains 200 hours of classical solo piano recordings. For ease of use, all tracks are resampled to 22.05 kHz.
For sampling, we use the second-order stochastic sampler from \cite{karras2022elucidating} (Algorithm 2), with the stochasticity parameter $S_\text{churn}=5$.
Also following \cite{karras2022elucidating}, we assign the noise variance as $\sigma_t=t$, with $t$ distributed according to: 
\begin{equation}\label{schedule}
    \tau_{i<T}=\left(\sigma_{\text{max}}^{\;\;\;\frac{1}{\rho}} 
    + \tfrac{i}{T-1}\left(
    \sigma_\text{min}^{\;\;\;\frac{1}{\rho}}
    -\sigma_\text{max}^{\;\;\;\frac{1}{\rho}}
    \right)\right)^\rho,
\end{equation}
where $T$ denotes the number of discretized steps and $i$ is the discretization index.
In these experiments
, we use $\sigma_\text{min}=10^{-4}$,  $\sigma_\text{max}=1$, and $\rho=13$. We found this schedule to be empirically effective for sampling with a reduced number of discretization steps. We use $T=35$ discretization steps for the bandwidth extension and inpainting experiments, while, for declipping, we raised the number of steps to $T=140$. Regarding the inference speed, with $T=35$ steps, CQT-Diff requires 17\,s for processing a 2.97-s segment in a GPU. The inference time augments up to 84\,s when reconstruction guidance is used, as this feature requires gradient computation.


The neural network $F_\theta^\prime$ was trained using the Adam optimizer with a learning rate of $2\cdot10^{-4}$. During training, the diffusion time $t$ is randomly sampled following \eqref{schedule}, but the parameters are set as $\sigma_\text{min}=10^{-6}$,  $\sigma_\text{max}=10$, and $\rho=10$\footnote{The distribution of noise levels used during includes a wider range of values than the ones used required during sampling. This allowed for flexibility during the exploration of the optimal sampling parameters without retraining.}. We set a segment size of 2.97 s ($2^{16}$ samples) and a batch size of 4, but, as the architecture is fully convolutional, the segment size can be freely modified during inference. The model was trained for 320k iterations, which took about three days on a single NVIDIA A100 GPU.


\subsection{Bandwidth Extension}\label{sec:bwe}
The first set of experiments concerns audio bandwidth extension, where the degradation function is a lowpass filter $\bm{y}=\mathcal{A}(\bm{x})=\text{LPF}(\bm{x})$. In contrast to previous works \cite{moliner2022behm, han2022nu, sulun2020filter}, our method does not suffer from filter generalization as a consequence of its fully generative nature, and, as we demonstrate in the companion webpage, can be applied after many different types of lowpass filters, even when the observations contain Gaussian noise. 
Nevertheless, we restrict the evaluation to FIR filters designed with a Kaiser window, with order 500 and two cutoff frequencies ($f_\text{c} =$ 1 and 3 kHz). In this setting, it is possible to apply \textit{data consistency} steps, where the observed low frequencies are replaced with the denoised estimate following 
 $\bar{\bm{x}}_0 = \bm{y} + \hat{\bm{x}}_0 - \text{LPF}(\hat{\bm{x}}_0)$, in a way similar to \cite{choi2021ilvr}. 

We conduct an objective evaluation using 53 complete recordings (about 6 h of music) from the MAESTRO test set,
which were not used during training.  
As objective metrics, we report in Table \ref{tab:objective_metrics_bwe} the averaged Log-Spectral-Distance (LSD), Frechet Audio Distance (FAD), and the F1 score (onset) of a transformer-based piano transcription model \cite{hawthorne2021sequence} (F1)\footnote{The MAESTRO dataset includes the ground-truth MIDI transcriptions.}.
 For this task, we experiment with applying our method (CQT-Diff) in two conditioning settings: using only \textit{data consistency} (DC) and  combining it with \textit{reconstruction guidance} (DC+RG), with the scaling parameter $\xi^\prime=0.1$. 
We include two conditions where the CQT-based neural architecture is replaced with one using the STFT (STFT-Diff) and raw audio (SaShiMi \cite{goel2022sashimi}). We also compare with BEHM-GAN \cite{moliner2022behm}, a recent method for music bandwidth extension which we retrained using the same dataset. The BEHM-GAN was trained using lowpass filters with randomized cutoff frequencies at around 3 kHz, so it could only be used as the baseline in the  $f_\text{c} =$ 3 kHz setting. Table \ref{tab:objective_metrics_bwe} also reports the number of trainable parameters (\#prms) as an indicator of the model capacity. 


    
\begin{figure}[t]
    \centering
    \includegraphics[width=0.93\columnwidth]{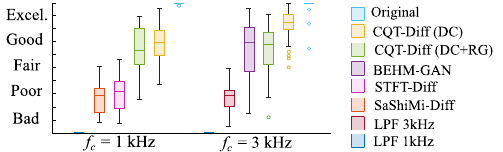}
    \caption{Results of the bandwidth extension listening test.}
    \label{boxplot_bwe}
\end{figure}

We conducted a subjective listening test following the MUSHRA recommendation.
A total of 12 volunteers were asked to rate the different methods in terms of perceptual audio quality, in comparison with a reference. The test was split into two parts, each evaluating the bandwidth extension from the cutoff frequencies of 1 kHz and 3 kHz.
Each part of the test contained 4 different audio examples.
Regarding the conditions, both parts shared the same low anchor (a lowpass-filtered sample at $f_\text{c}$ = 1 kHz), but the other conditions differed. In the 1 kHz part, we included the STFT-Diff and SaShiMi-Diff conditions, as well as the two variants of our method. In the 3 KHz part, the lowpass-filtered sample at $f_\text{c}$ = 3 kHz was included, as well as the BEHM-GAN baseline, but the STFT-Diff and SaShiMi-Diff conditions were excluded for length reasons.  The results of the subjective experiments are presented in Fig.~\ref{boxplot_bwe}.

After analyzing the results, we reached the following conclusions. 
(i) The CQT data representation exhibits a more favourable performance against the STFT or SaShiMi, as its MUSHRA and F1 scores are higher. 
    (ii) The use of RG is not critical for this task, as the (DC) and (DC+RG) conditions obtained comparable results. The reason may be that the architecture is explicitly designed to capture long-range data dependencies in the frequency range, thanks to the dilated convolutions.
    (iii) In the $f_\text{c} = 3$ kHz setting, the CQT-Diff (DC) condition obtained significantly better subjective scores than the baseline BEHM-GAN (p-value 1.4$\cdot10^{-5}$ in a paired t-test).


\begin{table}[t]
\caption{Objective metrics for bandwidth extension. The best result in each column is highlighted.}
\centering
\resizebox{1\columnwidth}{!}{%
\begin{tabular}{@{}l|lll|lll|l@{}}
\toprule
                                & \multicolumn{3}{c|}{$f_\text{c}$ = 1\;kHz}               & \multicolumn{3}{c|}{$f_\text{c}$ = 3\;kHz}  &              \\ 
\textbf{Method}                                & LSD $\downarrow$    & FAD $\downarrow$              & F1  $\uparrow$          & LSD $\downarrow$   & FAD $\downarrow$        & F1 $\uparrow$                & \#prms \\ \midrule
LPF                             & 1.29 & 12.47             & 0.67      & 0.81 & 3.50 & 0.73             &  -  \\
BEHM-GAN                        & -      & -                 & -            & 0.68 & 0.37  & 0.93       & 82M \\
STFT-Diff                       & 0.95 & \textbf{0.48} & 0.68 & 0.71 & 0.21  & 0.93 & 65M \\
SaShiMi-Diff                    & 0.97 & 4.06          & 0.63     & 0.70 & 1.37    & 0.93        & 19M   \\
CQT-Diff (DC)          & \textbf{0.86} & 0.51         & 0.79     & 0.66 & 0.22  & \textbf{0.94}         & 15M  \\
CQT-Diff (DC+RG)        & \textbf{0.86} & 0.50          & \textbf{0.80}    & \textbf{0.65} & \textbf{0.17} & \textbf{0.94}       & 15M   \\ \bottomrule
\end{tabular}
}
\label{tab:objective_metrics_bwe}
\end{table}

\subsection{Long Audio Inpainting}
We apply our framework for the task of long audio inpainting, where the degradation function is a linear binary mask 
$\mathcal{A}(\bm{x})=(1-\mathbbm{1}_{[t_\text{start}, t_\text{end}]})\bm{x}$, which deletes a long signal segment between $t_\text{start}$ and $t_\text{end}$.
For this task, the conditioning can be applied by using the DC steps as  
$\bar{\bm{x}}_0 = \bm{y} + \mathbbm{1}_{[t_\text{start}, t_\text{end}]}\hat{\bm{x}}_0$,
and/or using reconstruction guidance, directly applying \eqref{recguid}. 

We have evaluated the success of our method in long audio inpainting by means of a subjective experiment, which consisted of a multi-stimulus listening test. In the test, an 11-s excerpt of piano music with a long gap, i.e., silence, in the middle was presented to the participants as the reference. The participants were asked to rate the rest of the different conditions in terms of plausibility of the reconstruction. The test included four audio examples with three different gap lengths, forming a total of 12 pages of comparisons. We evaluate CQT-Diff in the same conditioning settings as in Sec. \ref{sec:bwe}, relying solely on the DC and combining it with RG, using the scaling parameter $\xi^\prime=0.35$. We compare our method against two recent baselines: GACELA \cite{marafioti2020gacela} and Catch-A-Waveform \cite{greshler2021catch} (CAW). 






\begin{figure}[t]
    \centering
    \includegraphics[width=0.93\columnwidth]{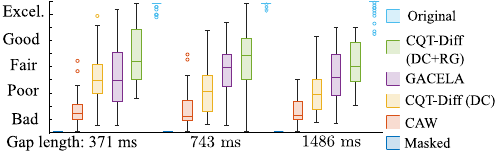}
    \caption{Results of the audio inpainting listening test.}
    \label{fig:results_inpainting}
\end{figure}
The main conclusions from the results of the
subjective experiment reported in Fig.~\ref{fig:results_inpainting} are the following:
(i) In contrast with the outcomes from Sec.~\ref{sec:bwe}, reconstruction guidance is critical for inpainting, given the significant difference in the results. Our explanation is that the neural architecture $F_\theta$ has not been explicitly designed to account for long-term time dependencies and, thus, the guidance term is necessary to ensure coherence. 
(ii) Although the confidence intervals overlap, the MUSHRA scores of CQT-Diff are marginally higher than GACELA. We conclude that the proposed method shows at least comparable performance against the baseline while being more flexible.

\subsection{Audio Declipping}

Audio declipping refers to canceling the distortion in an audio signal that has been hard-clipped. We treat declipping as a nonlinear inverse problem, where the degradation function is a saturating nonlinearity $\mathcal{A}(\bm{x}_0)=(|\bm{x}_0+c|-|\bm{x}_0-c|)/2$, where $c$ is the clipping threshold. In this case, we apply the conditioning on the clipped observations based on, uniquely, the RG strategy. 

We conducted an objective evaluation where CQT-Diff was compared against two popular audio declipping methods:  Analysis SPArse DEclipper (A-SPADE) \cite{kitic2015sparsity} and Social Sparsity with Persistent Empirical Wiener (SS-PEW) \cite{siedenburg2014audio}.
We report three objective metrics: FAD \cite{kilgour2019frechet}, PEAQ, and PEMO-Q.
The test set consists of 149 segments from different piano recordings of the MAESTRO test set, each of 2.97 s. Table \ref{tab:declipping} shows the evaluation results on two Signal-to-Distortion Ratio (SDR) settings: 1 dB and 10 dB.

\begin{table}[t]
\caption{Objective metrics on audio declipping tests.}
\label{tab:declipping}
\centering
\resizebox{1\columnwidth}{!}{%
\begin{tabular}{l|lll|lll}
\toprule
         & \multicolumn{3}{c|}{SDR = 1\;dB} & \multicolumn{3}{c}{SDR = 10\;dB} \\ 
         & FAD $\downarrow$   & PEAQ $\downarrow$    & PEMO-Q $\downarrow$   & FAD $\downarrow$     & PEAQ $\downarrow$     & PEMO-Q $\downarrow$   \\\midrule
Clipped  & 23.09   & -3.88   & -3.82   & 7.85    & -3.87    & -3.78   \\
CQT-Diff (RG) & \textbf{1.84}    & \textbf{-3.53}   & \textbf{-3.36}      & 0.68    & -2.99    &  -2.56       \\
A-SPADE   & 8.62    & -3.87   & -3.58   & 0.59    & -2.97    & -2.60   \\
SS-PEW    & 9.69    & -3.78   & -3.62   & \textbf{0.46}    & \textbf{-2.05}    & \textbf{-0.93}   \\ \bottomrule
\end{tabular}%
}
\end{table}


Thanks to the generative priors, CQT-Diff has an advantage over the other approaches in the most severe circumstances: when SDR = 1\;dB, it achieved higher scores than the compared baselines in the three reported metrics in  Table \ref{tab:declipping}. However, in moderate cases (SDR 
= 10 dB), the performance of CQT-Diff is upper-bounded by the still imperfect generation quality of the diffusion model. This explains why CQT-Diff is evidently outperformed by SS-PEW when SDR = 10\;dB, although having comparable performance to A-SPADE. 



%


\section{Conclusions}
We proposed CQT-Diff, a diffusion-based generative model which can be used as a flexible framework for audio restoration. 
In objective and subjective experiments, CQT-Diff excels in the task of audio bandwidth extension and shows competitive performance to the compared baselines in long audio inpainting and declipping, without the need of retraining for each task in particular. 
Apart from the three evaluated tasks, we emphasize that the proposed model could be flexibly applied for solving a much wider range of linear or nonlinear ill-posed problems, as long as the degradation is known.
The presented model is designed to benefit from a CQT that leverages data structure as an inductive bias.
However, we must admit that the benefits of the CQT are a conjecture that is not empirically substantiated in the current paper and should to be tackled in future work
Future work should also target adapting the model to be trained in a large-scale data regime extending beyond piano music, which would allow for general audio restoration applications. 
This work shows the potential of diffusion models for audio restoration, and how they can benefit from using invertible time-frequency representations. 
\section{Acknowledgments}
We thank Tero Karras, Miika Aittala, Samuli Laine, and Timo Aila for the helpful discussions and for proposing the idea of using a Constant-Q Transform. We acknowledge the computational resources provided by the Aalto Science-IT project.

\bibliographystyle{IEEEbib}
\bibliography{refsICASSP}

\end{document}